\documentstyle[12pt]{article}
\begin{document}
\baselineskip=20pt

\title{Quantum fields in anti de Sitter spacetime 
and degrees of freedom in the bulk/boundary
correspondence
}

\author{\large Henrique Boschi-Filho\footnote{\noindent e-mail: 
boschi @ if.ufrj.br}\,  
and 
Nelson R. F. Braga\footnote{\noindent e-mail: braga @ if.ufrj.br}
\\ 
\\ 
\it Instituto de F\'\i sica, Universidade Federal do Rio de Janeiro\\
\it Caixa Postal 68528, 21945-970  Rio de Janeiro, RJ, Brazil}
 
\date{}

\maketitle

\vskip 3cm

\begin{abstract} The quantization of a scalar field in anti de Sitter 
spacetime using Poincar\'e coordinates is considered. 
We find a discrete spectrum that is consistent with a possible
mapping between bulk and boundary quantum states.

\end{abstract}


\vfill\eject


The holographic principle asserts that the degrees of freedom of a 
quantum system with gravity can be represented by a theory on the 
boundary\cite{HOL1,HOL2,HOL3}.
The presence of gravity makes it possible to define a mapping 
between  theories defined in 
manifolds of different dimensionality. 
One interesting realization of the holographic principle can be done
in a space of constant negative curvature, the anti de Sitter ($AdS$) 
spacetime.
Such a realization was proposed by Maldacena in the form of a 
conjecture\cite{Malda} on the equivalence (or duality) of the large 
$N$ limit of $SU(N)$ superconformal field theories in $n$
dimensions and supergravity on anti de Sitter spacetime in $n+1$ 
dimensions $(AdS/CFT\,$ correspondence). 
Then, using Poincar\'e coordinates in the $AdS$ bulk,
 Gubser, Klebanov and Polyakov \cite{GKP} and  Witten \cite{Wi} 
found prescriptions for relating theories that live in the bulk and 
on the boundary, where the AdS solutions play the role of classical 
sources for the boundary field correlators.

\bigskip

Despite the fact that field quantization in AdS in terms of global 
coordinates has been known for a long time \cite{QAdS1,QAdS2}, 
the corresponding formulation in Poincar\'e coordinates and 
thus a comprehensive picture of holography in terms of bulk
quantum fields is still lacking.
The aim of the present letter is to investigate a quantum theory 
for a scalar field in the AdS bulk in terms of Poincar\'e coordinates. 
We will see that the dimensionality of the phase space is such that 
a mapping between this theory and states on the boundary is possible. 
This conclusion essentially depends on the fact that the
AdS space in Poincar\'e coordinates should be compactified, 
as it happens in the usual global coordinates, in order to include 
appropriate boundary conditions at infinity and find a 
consistent quantization. 
Although the Poincar\'e coordinates extend to infinity
we will need to introduce a finite radius (cutoff) $R$ 
corresponding to the fact that we can not 
represent the whole compactified AdS space used in the AdS/CFT 
correspondence (including the infinity) into just one single set 
of Poincar\'e coordinates. 
Naturally, we can take $R$ large enough to describe as much of
the entire AdS space as we want.
We are going to see in the following that this result is in agreement 
with the counting of degrees of freedom in the bulk/boundary correspondence.


The anti-de Sitter spacetime of $\,n+1$ dimensions can be represented 
as the hyperboloid ($\Lambda\,=\,$constant)

\begin{equation}
X_0^2 + X_{n+1}^2 - \sum_{i=1}^n X_i^2\,=\,\Lambda^2
\end{equation}

\noindent in a flat $n+2$ dimensional space with metric
\begin{equation}
ds^2\,=\, - d X_0^2 - dX_{n+1}^2 + \sum_{i=1}^n dX_i^2.
\end{equation}

The  so called global coordinates $\,\rho,\tau,\Omega_i\,$ for 
$AdS_{n+1}\,$  can be defined by \cite{Malda2,Pe}

\begin{eqnarray}
\label{global}
X_0 &=& \Lambda \,\sec\rho\, \cos \tau \nonumber\\
X_i &=& \Lambda \,\tan \rho\, \,\Omega_i\,\,\,\,\,\,\,\,
(\,\sum_{i=1}^n \,\Omega^2_i\,=\,1\,) \nonumber\\
X_{n+1} &=& \Lambda \sec \rho \,\sin\tau \,,
\end{eqnarray}

\noindent with ranges $0\le \rho <\pi/2$ and $0\le\tau< 2\pi\,$.

Poincar\'e coordinates $\,z \,,\,\vec x\,,\,t\,$ can be introduced by
\begin{eqnarray}
\label{Poincare}
X_0 &=& {1\over 2z}\,\Big( \,z^2\,+\,\Lambda^2\,
+\,{\vec x}^2\,-\,t^2\,\Big)
\nonumber\\
X_i &=& {\Lambda x^i \over z}
\nonumber\\
X_n &=& - {1\over 2z}\,
\Big( \,z^2\,-\,\Lambda^2\,+\,{\vec x}^2\,-\,t^2\,\Big)
\nonumber\\
X_{n+1} &=& {\Lambda t \over z}\,,
\end{eqnarray}

\noindent where $\vec x $ has $n-1$ components and 
 $0 \le z < \infty $. In this case the $\,AdS_{n+1}\,$ measure with 
Lorentzian signature reads

\begin{equation}
\label{metric}
ds^2=\frac {\Lambda^2 }{( z )^2}\Big( dz^2 \,+(d\vec x)^2\,
- dt^2 \,\Big)\,.
 \end{equation}

Then the $AdS$ boundary described by usual Minkowski 
coordinates $\vec x$ , $t$  corresponds to the 
region $\,z\,=\,0\,$ plus a ``point'' at infinity 
($z\,\rightarrow\,\infty\,$).

In order to gain some insight into the form of the spectrum associated 
with quantum fields to be defined in the AdS/CFT framework, 
let us discuss an essential point of the correspondence: 
the mapping between the degrees of freedom of the bulk volume and those
of  the boundary hypersurface.  
The metric is singular at $z = 0$, so the prescriptions \cite{GKP,Wi} 
for calculating field correlators should be first taken at some small 
$z$ that then goes to zero \cite{MV,FMMR,Pe,BB}. 
In the same way, we will consider the boundary 
to be at some small $\,z\,=\,\delta\,$.

So, let us consider at $\,z\,=\,\delta\,$  a hypersurface of area 
$\,\Delta A \,$ corresponding to variations 
$\,\Delta x^1 \, ... \, \Delta x^{n-1} \,$ in the space coordinates:
\begin{equation}
\,\Delta A \,=
\left(\frac{\Lambda}{\delta}\right)^{n-1}
\,\Delta x^1 \, ... \, \Delta x^{n-1} \,
\end{equation}
 
\noindent 
and calculate the volume generated by this surface moving $z$ from 
$\delta$ to $\infty$, finding
\begin{equation}
\Delta V \,=\,\Lambda \,{\Delta A \over n-1}.
\end{equation}

\noindent This is the expected result that the volume is proportional 
to the area in the bulk/boundary correspondence for a fixed $\Lambda$. 
In order to count the degrees of freedom we can split 
$\Delta V$ in $\ell$ pieces of equal volume corresponding to cells whose 
boundaries are hypersurfaces located at
\begin{equation}
z_j\,=\, {\delta\over \sqrt[n-1]{1- j/\ell}}\,\,\,\,,
\end{equation}

\noindent with $\,j\,=\,1,..,\ell -1\,$. Note that the last cell extends 
to infinity. 
These volume cells can be mapped into the area $\,\Delta A\,$ 
by dividing it also in $\ell$ parts. This way, one finds a one to one
mapping between degrees of freedom of bulk and boundary.  
This analysis shows us that despite the fact that the variable $z$ 
has an infinite range, the volume, and thus the associated degrees of 
freedom, corresponding to a finite surface, are finite. 
One could, in a 
simplified way, think of the system as being "in a box" in terms of 
degrees of freedom, with respect to $z$.
We could have changed $z$ to a variable that measures the volume,
say $\,\zeta\,=\, 1/\delta^{n-1}\,- \, 1/z^{n-1} \,$, to explicitly find 
this compact role of the radial coordinate $z$ but this would not be 
better than just going to global coordinates, eqs.(\ref{global}). 
However we want to see from the point of view of the 
Poincar\'e coordinates, where the AdS/CFT correspondence takes its 
more natural form, how does this compactified character of the radial 
coordinate manifests itself. 

Then, let us  consider a massive scalar field $\phi$ in the $\,AdS_{n+1}\,$
spacetime described by Poincar\'e coordinates with action

\begin{equation}
\label{action1}
I[\phi ]\,=\, {1\over 2} \int d^{n+1}x \sqrt{g}\,
\left(\partial_\mu \phi \, \partial^\mu \phi
+m^2\,\phi^2 \right)
\,\,.
\end{equation}
  
\noindent where we take $x^0\,\equiv\,z\,,\,x^{n+1}\,\equiv\,t\,$,
$\sqrt{g}\,=\,(x^0)^{-n-1}\,$ and $\mu\,=\,0,1,...,n+1\,$.

The classical equation of motion reads                                                                                                                                                                                                                                                                                                                                                                                                          

\begin{equation}
\label{motion}
\left(\nabla_\mu \nabla^\mu - m^2\right) \phi\,
=\,{1\over \sqrt{g}} 
\partial_\mu 
\Big( \, \sqrt{g} \partial^\mu \phi \,\Big) 
- m^2\phi\,=\,0\,\,
\end{equation}

\noindent and the solutions can be found \cite{BKL,BKLT} in terms of Bessel
functions using the ansatz 
$\phi\,=\, e^{-i\omega t\,+\,i\vec k \cdot \vec x} z^{n/2} \chi(z)$. 
Taking $\omega^2\,>{\vec k}^2$ and defining 
$u\,=\,\sqrt{ \omega^2\,-\,{\vec k}^2\,}\,$ 
and $\nu=\frac 12\sqrt{n^2+4m^2}$,
we have two independent solutions
\begin{equation}
\Phi^{\pm}\,=\, e^{-i\omega t\,+\,i\vec k \cdot \vec x} z^{n/2} 
J_{\pm\nu}(uz),
\end{equation}

\noindent if $\nu$ is not integer. 
If $\nu$ is integer one can take $\Phi^+\,$ and 
\begin{equation}
\Phi^-\,=\, e^{-i\omega t\,+\,i\vec k \cdot \vec x} z^{n/2} Y_{\nu}(uz)
\end{equation}


\noindent as independent solutions. 

On the other hand, if ${\vec k}^2>\omega^2$ the solution is 
\begin{equation}
{\overline \Phi}\,=\, e^{-i\omega t\,+\,i\vec k \cdot \vec x} z^{n/2}
K_{\nu}(qz)
\end{equation}

\noindent where $q\,=\,\sqrt{ {\vec k}^2\,-\,\omega^2\,}\,$ 
(the second solution in this case is proportional to $I_\nu(qz)$ 
which is  divergent as $z\to\infty$).

As discussed in refs.\cite{BKL,BKLT}, $\,\Phi^+ \,$ are the only 
normalizable solutions in the range 
$0\,<\, z\,<\,\infty\,$. They are thus the natural candidates 
for the role of quantum fields, if we want to be able to take the limit
of $\delta\,\rightarrow\,0$ at the end.
One could then naively think of just adding all possible solutions 
$\Phi^+ \,$ and thus building up a quantum field like

\begin{equation}
\,\int du\,d^{n-1}k \, f(\vec k , u) 
\Phi^+ (\vec k, u)\,\,+ c.c.\,\,.
\end{equation}

\noindent However, from our previous analysis of degrees of freedom, we 
expect to find a discrete spectrum associated with the radial coordinate 
$z$. Such a discretization would be in accordance with the results coming 
from the quantization in global coordinates\cite{QAdS1,QAdS2,BKL,BKLT}.

One can understand why this discretization also takes place in Poincare 
coordinates by considering a simpler situation:
the stereographic mapping of the surface of  a sphere on a plane.
One can map the points of a sphere on a plane plus a point at infinity.
However looking at the sphere one sees that this compact manifold has 
discrete sets of eigenfunctions but looking at the plane how can we 
realize that the spectrum of eigenfunctions in the radial direction 
would be discrete?
In close analogy with the case of finite volumes $\Delta V$ for 
$z\,\rightarrow\,\infty\,$ in AdS discussed above, here in the case of 
the sphere if we calculate 
the area of the plane taking the metric induced by the sphere into account
we would find a finite value (equal to the area of the sphere).
So, the radial coordinate on the plane  looks also like a "compact" one
(in the sense of degrees of freedom or whatever we associate with area 
cells) in the same way as the $z$ coordinate of AdS.
The extra point at infinity corresponds to the fact that we should  
impose the condition that going to infinity in any direction would led to 
the same point. This condition 
would mean that either the functions on the plane have no angular 
dependence or they vanish as the radial coordinate tend to zero. 
These conditions would not yet lead to a discrete spectrum (in the radial 
direction).  This problem is simply related to the fact 
that the point at infinity, which has zero measure,  is not represented
on the plane.
If instead of using just one plane, we project the sphere on two 
different planes
we would be able to represent all the points of the sphere.
We could even choose one of the mappings to cover "as much of the surface 
of the sphere" as we want. As long as we map it into two disks of finite 
radius (with an appropriate matching boundary 
condition), one would then clearly see that the spectrum of eigenfunctions 
is discrete.

Now coming back to the AdS case, a consistent quantization  
in this space in global coordinates 
\cite{QAdS1,QAdS2} requires the introduction of 
boundary conditions at the surface corresponding to $\rho\,=\,\pi/2$
in order to have a well 
defined Cauchy problem. So, one must consider a compactified AdS 
including $\rho\,=\,\pi/2$ in order to find a consistent 
theory. 

The  limit $\,z\,\rightarrow\,\infty\,$ in Poincar\'e coordinates
(\ref{Poincare}) corresponds to a point that in global coordinates 
sits in the hypersurface $\rho\,=\,\pi/2$. 
Thus, this hypersurface is not 
completely represented in just one set of Poincar\'e coordinates. 
In the same way as in the case of the sphere, we can solve this problem
mapping the compactified 
AdS in two sets of Poincar\'e coordinates. 
We can simply stop at  $z\,=\,R\,$ in one set 
and map the rest, including the point at infinity, in  a second set.
We can take $R$ arbitrarily large so that we can map as much of the 
compactified AdS spacetime as we want in just one set. 

In this region $0\,\le\,z\,\le R\,$ we can introduce as quantum fields
\begin{equation}
\label{QF}
\Phi(z,\vec x,t)\,=\,\sum_{p=1}^\infty \,
\int {d^{n-1}k \over (2\pi)^{n-1}}\,
{z^{n/2} \,J_\nu (u_p z ) \over R w_p(\vec k ) 
\,J_{\nu\,+\,1} (u_p R ) }
\lbrace { a_p(\vec k )\ }
 e^{-iw_p(\vec k ) t +i\vec k \cdot \vec x}\,
\,+\,\,c.c.\rbrace
\end{equation}

\noindent where $w_p(\vec k ) \,=\,\sqrt{ u_p^2\,+\,{\vec k}^2}$
 and $u_p$ are such that $J_\nu(u_p R)=0$.

Imposing that the operators $a,a^{\dagger}\,$ satisfy the commutation 
relations

\begin{eqnarray}
\Big[ a_p(\vec k )\,,\,a^\dagger_{p^\prime}({\vec k}^\prime  )
\Big]&=& 2\, (2\pi)^{n-1} w_p(\vec k )   
\delta_{p\,  p^\prime}\,\delta^{n-1} (\vec k -
{\vec k}^\prime )  \nonumber\\
\Big[ a_p(\vec k )\,,\,a_{p^\prime}({\vec k}^\prime  )
\Big] &=& \Big[ a^\dagger_p(\vec k )\,,\,
a^\dagger_{p^\prime}({\vec k}^\prime  ) \Big]\,=\,0
\end{eqnarray}

\noindent we find, for example, for the equal time commutator of field
 and time derivative

\begin{equation}
\Big[ \Phi (z,\vec x ,t)\,,\,{\partial\Phi\over \partial t}(z^\prime,
\vec x^\prime ,t)\,\Big]\,=\,i z^{n-1} \delta (z - z^\prime) 
\delta (\vec x - {\vec x}^\prime )\,.
\end{equation}

Now considering  again the field (\ref{QF}) we realize that the 
discretization of the spectrum makes it possible to map the phase space
$\,u_p\,,\,\vec k \,$ into the momentum space of a field theory 
defined on the boundary, in the same way as we can map an infinite
but enumerable set of lines into just one line.

Taking, for simplicity, $AdS_3$ where $\vec k$ has just one component,
the phase space in the bulk would be an enumerable set of lines each
one corresponding to the continuous values of $-\infty <k<\infty $ 
and one fixed value of $p$.
One can map these lines into just one line, corresponding to some 
momentum, say $\kappa_b\,$ on the boundary, by dividing the line of 
$\kappa_b\,$ into segments of finite size.
This kind of mapping would not be possible if the spectrum were not 
discrete,
as one can not define a one to one mapping between a plane and a line.
So, the discretization of the spectrum is a necessary ingredient
for the holographic mapping to hold.

In conclusion, we have obtained a quantum scalar field in the AdS bulk
that exhibits a discrete spectrum associated with the radial Poincare 
coordinate. This result was obtained taking into account the 
compactification of AdS that in Poincare coordinates  corresponds 
to adding a point at infinity. 
This discretization is in agreement with the counting of degrees of freedom
suggested by the holographic principle. However it is in contrast to the 
continuous spectrum found in ref.\cite{Cald}.

\section*{Acknowledgments} 
The authors were partially supported by CNPq, FINEP and FUJB 
- Brazilian research agencies. We also thank  Mauricio Calv\~ao,
Juan Mignaco, Cassio Sigaud and Arvind Vaidya for interesting discussions.




\begin{thebibliography}{30}

\bibitem{HOL1} G. 't Hooft, "Dimensional reduction in quantum gravity"
in Salam Festschrifft, eds. A. Aly, J. Ellis and S. Randjbar-Daemi,
 World Scientific, Singapore, 1993, gr-qc/9310026.

\bibitem{HOL2} L. Susskind, J. Math. Phys. 36 (1995) 6377.

\bibitem{HOL3} L. Susskind and E. Witten, "The holographic bound in anti-de 
Sitter space", SU-ITP-98-39, IASSNS-HEP-98-44, hep-th 9805114.

\bibitem{Malda} J. Maldacena, Adv. Theor. Math. Phys. 2 (1998) 231.

\bibitem{GKP} S. S. Gubser , I.R. Klebanov and A.M. Polyakov, 
Phys. Lett. B428 (1998) 105.

\bibitem{Wi} E. Witten, Adv. Theor. Math. Phys. 2 (1998) 253.

\bibitem{QAdS1} S. J. Avis, C. J. Isham and D. Storey, Phys. Rev. D18
(1978) 3565.

\bibitem{QAdS2} P. Breitenlohner and D. Z. Freedman, 
Phys. Lett. B115(1982) 197; Ann. Phys. 144 (1982) 249.


\bibitem{Malda2} O. Aharony, S.S. Gubser, J. Maldacena, 
H. Ooguri and Y. Oz, Phys. Rept 323 (2000) 183.

\bibitem{Pe} J. L. Petersen, Int. J. Mod. Phys. A14 (1999) 3597.

\bibitem{MV} W. Mueck and K. S. Viswanathan, Phys. Rev. D58(1998)041901.

\bibitem{FMMR} D. Z. Freedman, S. D. Mathur, A. Matusis and L. Rastelli,
Nucl. Phys. B546 (1999) 96.
 
\bibitem{BB} H. Boschi-Filho and N. R. F. Braga, Phys. Lett. B 
471 (1999) 162.


\bibitem{BKL} V. Balasubramanian, P. Kraus and A. Lawrence, 
Phys. Rev. D59 (1999) 046003;

\bibitem{BKLT}
V. Balasubramanian, P. Kraus, A. Lawrence and S. P. Trivedi, 
Phys. Rev. D59 (1999) 1046021;

\bibitem{Cald}
M. M. Caldarelli, Nucl. Phys. B549 (1999) 499.

\end{thebibliography}
\end{document}